
\input phyzzx
\PHYSREV
\hoffset=1.25in
\hfuzz=8pt

\def\Re{{\cal R \mskip-4mu \lower.1ex \hbox{\it e}\,}}
\def\Im{{\cal I \mskip-5mu \lower.1ex \hbox{\it m}\,}}

\def\etal{{\it et al.}}

\def\sub#1{_{\lower.25ex\hbox{$\scriptstyle#1$}}}
\def\sul#1{_{\kern-.1em#1}}
\def\sll#1{_{\kern-.2em#1}}
\def\sbl#1{_{\kern-.1em\lower.25ex\hbox{$\scriptstyle#1$}}}
\def\ssb#1{_{\lower.25ex\hbox{$\scriptscriptstyle#1$}}}
\def\sbb#1{_{\lower.4ex\hbox{$\scriptstyle#1$}}}

\def\tev{\,{\rm TeV}}

\def\to{\rightarrow}
\def\mh{\ifmmode m\sbl H \else $m\sbl H$\fi}
\def\mch{\ifmmode m_{H^\pm} \else $m_{H^\pm}$\fi}
\def\mt{\ifmmode m_t\else $m_t$\fi}
\def\mc{\ifmmode m_c\else $m_c$\fi}
\def\mz{\ifmmode M_Z\else $M_Z$\fi}
\def\mw{\ifmmode M_W\else $M_W$\fi}
\def\mws{\ifmmode M_W^2 \else $M_W^2$\fi}
\def\mhs{\ifmmode m_H^2 \else $m_H^2$\fi}
\def\mzs{\ifmmode M_Z^2 \else $M_Z^2$\fi}
\def\mts{\ifmmode m_t^2 \else $m_t^2$\fi}
\def\mcs{\ifmmode m_c^2 \else $m_c^2$\fi}
\def\mchs{\ifmmode m_{H^\pm}^2 \else $m_{H^\pm}^2$\fi}
\def\ztwo{\ifmmode Z_2\else $Z_2$\fi}
\def\zone{\ifmmode Z_1\else $Z_1$\fi}
\def\mtwo{\ifmmode M_2\else $M_2$\fi}
\def\mone{\ifmmode M_1\else $M_1$\fi}
\def\tb{\ifmmode \tan\beta \else $\tan\beta$\fi}
\def\xw{\ifmmode x\sub w\else $x\sub w$\fi}
\def\ch{\ifmmode H^\pm \else $H^\pm$\fi}
\def\lum{\ifmmode {\cal L}\else ${\cal L}$\fi}
\def\inpb{\ifmmode {\rm pb}^{-1}\else ${\rm pb}^{-1}$\fi}
\def\infb{\ifmmode {\rm fb}^{-1}\else ${\rm fb}^{-1}$\fi}
\def\epem{\ifmmode e^+e^-\else $e^+e^-$\fi}
\def\ppb{\ifmmode \bar pp\else $\bar pp$\fi}

\newskip\zatskip \zatskip=0pt plus0pt minus0pt
\def\matth{\mathsurround=0pt}
\def\lsim{\mathrel{\mathpalette\atversim<}}

\def\atversim#1#2{\lower0.7ex\vbox{\baselineskip\zatskip\lineskip\zatskip
  \lineskiplimit 0pt\ialign{$\matth#1\hfil##\hfil$\crcr#2\crcr\sim\crcr}}}

%

\catcode`@=12
\catcode`@=11
\def\p@bblock{\begingroup\tabskip=\hsize minus\hsize
   \baselineskip=1.5\ht\strutbox\topspace-2\baselineskip
   \halign to \hsize{\strut ##\hfil\tabskip=0pt\crcr
   \the\Pubnum\cr
   \the\date\cr \the\pubtype\cr}\endgroup}
\Pubnum{\bf ANL-HEP-PR-92-34}
\date={June 1992}
\pubtype={}
\titlepage
\hoffset=1.25in
\title{THE EFFECTS OF DETECTOR DESCOPING AND NEUTRAL BOSON MIXING ON
NEW GAUGE BOSON PHYSICS AT THE SSC
\footnote{*}{Work supported by the U.S. Department of
Energy, Division of High Energy Physics, Contracts W-31-109-ENG-38
and W-7405-Eng-82.} }

\author{ J.L.~Hewett,$^a$\footnote{\dag}{Research supported by an SSC
Fellowship
from the Texas National Research Laboratory Commission} \ and \
T.G.Rizzo$^{a,b}$}
\smallskip
\smallskip
\address{$^a$High Energy Physics Division, Argonne National Laboratory,
Argonne, IL \ 60439}
\smallskip
\address{$^b$ Ames Laboratory and Department of Physics,
Iowa State University,  Ames , IA \ 50011}

\vskip.25in
\abstract

 We examine how the abilities of an SDC-like detector to discover and identify
the origin of a new neutral gauge boson are affected by $Z_1-Z_2 $ mixing and
by variations in detector parameters such as lepton pair mass resolution,
particle identification efficiency, and rapidity coverage.  Also examined
is the sensitivity of these results to variations in structure function
uncertainties and uncertainties in the machine integrated luminosity.  Such
considerations are of importance when dealing with the issues of detector
descoping and design.

\endpage
\def\MPL #1 #2 #3 {Mod.\ Phys.\ Lett.\ {\bf#1},\ #2 (#3)}
\def\NPB #1 #2 #3 {Nucl.\ Phys.\ {\bf#1},\ #2 (#3)}
\def\PLB #1 #2 #3 {Phys.\ Lett.\ {\bf#1},\ #2 (#3)}
\def\PR #1 #2 #3 {Phys.\ Rep.\ {\bf#1},\ #2 (#3)}
\def\PRD #1 #2 #3 {Phys.\ Rev.\ {\bf#1},\ #2 (#3)}
\def\PRL #1 #2 #3 {Phys.\ Rev.\ Lett.\ {\bf#1},\ #2 (#3)}
\def\RMP #1 #2 #3 {Rev.\ Mod.\ Phys.\ {\bf#1},\ #2 (#3)}
\def\ZPC #1 #2 #3 {Z.\ Phys.\ {\bf#1},\ #2 (#3)}
\def\IJMP #1 #2 #3 {Int.\ J.\ Mod.\ Phys.\ {\bf #1},\ #2 (#3)}
\Ref\ssczp{J.L.\ Hewett and T.G.\ Rizzo, in {\it Proceedings of the 1988
Snowmass Summer Study on High Energy Physics in the 1990's}, Snowmass, CO
1988, ed.\ S.\ Jensen;
V.\ Barger \etal, \PRD D35 166 1987 ; L.S.\ Durkin and P.\ Langacker,
\PLB B166 436 1986 ; F.\ Del Aguila, M.\ Quiros, and F.\ Zwirner, \NPB B287
419 1987 , {\bf B284}, 530 (1987) ; J.A.\ Grifols, A.\ Mendez, and
R.M.\ Barnett, \PRD D40 3613 1989 ;  N.G.\ Deshpande and
J.\ Trampetic, \PLB B206 665 1988 ; N.G.\ Deshpande, J.F.\ Gunion, and
F.\ Zwirner in {\it Proceedings of the Workshop on Experiments,
Detectors, and Experimental Areas for the Supercollider}, Berkeley, CA, 1987,
ed.\ R.\ Donaldson and M.\ Gilchriese; F.\ del Aguila \etal,
\PLB B201 375 1988 , and {\bf B221}, 408 (1989);
S.\ Nandi, \IJMP A2 1161 1987 .}
\Ref\lhczp{P.\ Chiappetta \etal, to appear in the {\it Proceedings of the Large
Hadron Collider Workshop}, Aachen, Germany, 1990.}
\Ref\us{J.L.\ Hewett and T.G.\ Rizzo, \PRD D45 161 1992 .}
\Ref\bigref{J.L.\ Hewett and T.G.\ Rizzo, ANL Report ANL-HEP-PR-92-33 (1992);
M.\ Cveti\v c and P.\ Langacker, \PRD D46 R14 1992 , \PRD D42 1797 1990 ,
and Univ.\ Pennsylvania
Report UPR-514-T (1992); F.\ del Aguila \etal, Univ.\ Granada Report
UG-FT-22/92 (1992); M.\ Cveti\v c, B.\ Kayser, and P.\ Langacker, \PRL 68 2871
1992 ; H.\ Haber in {\it Proceedings of the 1984 Summer Study on the Design
and Utilization of the Superconducting Supercollider}, ed.\ R.\ Donaldson and
J.G.\ Morfin (1984); J.D.\ Anderson, M.H.\ Austern, and R.N.\ Cahn,
\PRD D46 290 1992 ; A.\ Fiandrino and P.\ Taxil, \PRD D44 3490 1991 ;
K.\ Whisnant, in {\it Proceedings of the 1990 Summer Study on High Energy
Physics - Research Directions for the Decade}, Snowmass, CO, 1990,
ed.\ E.\ Berger;  F.\ del Aguila and J.\ Vidal, \IJMP A4 4097 1989 ; B.\
Aveda \etal, in {\it Proceedings of the 1986 Summer Study on the Physics of the
Superconducting Supercollider}, Snowmass, CO, 1986 eds.\ R.\ Donaldson and
J.N.\ Marx.}
\Ref\sdc{For detailed descriptions of the SDC detector, see the
SDC Technical Design Report, E.\ Berger \etal, SDC Report SDC-92-201 (1992);
I.\ Hinchliffe, M.\ Mangano, and M.\ Shapiro, SDC Report SDC-90-00036 (1990);
I.\ Hinchliffe, SDC Report SDC-90-00100 (1990); I.\ Hinchliffe, M.\ Shapiro,
and J.L.\ Siegrist, SDC Report SDC-90-00115 (1990);
G.\ Eppley and H.E.\ Miettinen, SDC Reports SDC-90-00125 (1990),
and SDC-91-00009 (1991).}
\Ref\esix{For a review, see, J.L.\ Hewett and T.G.\ Rizzo, \PR 183 193 1989 .}
\Ref\pertsm{In the case of the Standard Model, see, B.W.\ Lee, C.\ Quigg, and
H.B.\ Thacker, \PRL 38 883 1977 ;
\PRD D16 1519 1977 ; M.S.\ Chanowitz, M.A.\ Furman, and I.\ Hinchliffe,
\NPB B153 402 1979 ; W.\ Marciano, G.\ Valencia, and S.\ Willenbrock, \PRD D40
1725 1989 .}
\Ref\pert{For earlier analyses of constraints on the exotic $E_6$ fermions,
see, R.W.\ Robinett, \PRD D34 182 1986 ;
 M.\ Dress and X.\ Tata, \PLB B196 65 1987 ; M.\ Dress, \NPB B298 333 1988 ;
D.\ London \etal, \PRD D37 799 1988 .}
\Ref\lrm{For a review and original references, R.N.\ Mohapatra, {\it
Unification and Supersymmetry}, (Springer, New York, 1986).}
\Ref\alrm{E.\ Ma, \PRD D36 274 1987 ; \MPL A3 319 1988 ; K.S.\ Babu \etal,
\PRD D36 878 1987 ; V.\ Barger and K.\ Whisnant, \IJMP A3 879 1988 ;
J.F.\ Gunion \etal, \IJMP A2 118 1987 ; T.G.\ Rizzo, \PLB B206 133 1988 .}
\Ref\ssm{The `sequential' standard model (SSM) contains a \ztwo\ which is just
a heavy version of the Standard Model $Z$-boson with identical couplings.}
\Ref\mtung{J.C.\ Morfin and W.-K.\ Tung, \ZPC C52 13 1991 .}
\Ref\xman{We thank X.\ Tata for bringing up this point.}
\Ref\lepzp{P.\ Langacker and M.\ Luo, \PRD D45 278 1992 ; P.\ Langacker,
M.\ Luo, and A.\ Mann, Rev.\ Mod.\ Phys.\ {\bf 64}, 87 (1992); F.\ del
Aguila, W.\ Hollik, J.M.\ Moreno, and M.\ Quiros, \NPB B372 3 1992 ;
E.\ Nardi, E.\ Roulet, and D.\ Tommasini, Univ.\ of Michigan Report
UM-TH-92-07 (1992).}

\endpage

\chapter{INTRODUCTION}

If a new neutral gauge boson (\ztwo) exists in the TeV mass range and has
couplings to both $q\bar q$ and $\epem$ at electroweak strength or larger, it
will be copiously produced and detected at hadron supercolliders such as the
SSC\refmark\ssczp\ and LHC\rlap.\refmark\lhczp\  Once a \ztwo\ is observed at
these colliders, the real challenge begins: determining the extended
electroweak model from which the \ztwo\ originated.  To meet this challenge,
all possible information about the couplings of the \ztwo\ must be
gathered\rlap,\refmark{\us,\bigref} and unfortunately,
hadron colliders provide few tools with which to work.  In our earlier
analysis\rlap,\refmark\us\ we began to address
these issues for a real SSC detector, the SDC\rlap.\refmark\sdc\  Specifically,
we examined the capability of the SDC to (i) directly determine the various
couplings of the \ztwo\ and (ii) determine the maximum value of  the \ztwo\
mass for which adequate statistical power is available to distinguish new
neutral gauge bosons from two different extended electroweak models.  The
latter is referred to as the ID-limit.  The measurable quantities used in this
analysis are the new gauge boson mass (\mtwo), the width ($\Gamma_2$), the
production cross section $(\sigma)$ for the reaction $pp\to Z_2\to
\ell^+\ell^-$, and the leptonic forward-backward asymmetry ($A_{FB}$) of the
\ztwo, folded together with the
anticipated SDC detector properties such as rapidity coverage, lepton-pair mass
resolution, and particle identification efficiency, as well as the luminosity
uncertainty of the SSC and the theoretical uncertainties due to our lack of
detailed knowledge of the parton distribution functions.

The purpose of the present work is to re-examine our previous results in order
to explore their sensitivity to possible variations in the capabilities of the
SDC detector, improvements in our knowledge of the parton densities and the
integrated machine luminosity, as well as to mixing between the \ztwo\
and the Standard Model (SM) $Z$-boson.  These considerations are particularly
relevant when dealing with issues of detector descoping and design.  We will
see below that, for a limited class of models, our previous conclusions could
be modified by as much as $\simeq 26\%$ from variations in the above detector
and machine characteristics, while the incorporation of neutral gauge boson
mixing does not significantly alter our results.  This
paper is organized such that we first examine the effects of detector
descoping,
neglecting gauge boson mixing, and then we investigate the contributions of
mixing, using a set of default detector parameters.  We refer the
interested reader to Ref.~\us\ for the full details of our analysis procedure.

Before discussing the main issues of this paper, we first briefly comment on
the influence of another assumption on our results; the omission of
possible contributions to $\Gamma_2$ arising from the existence of any new
particles not contained in the SM.  Most
extended electroweak models contain various exotic
particles into which the \ztwo\ may also decay.  For example, in $E_6$ theories
each generation lies in the {\bf 27} representation\rlap,\refmark\esix\ which
contains the standard fermions, a right-handed neutrino, and 11 additional
fields.  These additional fields are comprised of the following:
a color-triplet, iso-scalar, $Q=-1/3$ fermion denoted by $h$; a color singlet,
$Q=0,~{\rm and}~-1$, iso-doublet denoted by $N,~{\rm and}~E$, respectively, and
their conjugate fields; and a color singlet, iso-singlet, neutral fermion,
designated by $S^c$.  Most of these exotic fermions acquire their masses from
the same vacuum expectation value (vev) that generates the \ztwo\ mass,
and hence it is reasonable to expect that the exotics will have masses of the
same order as \mtwo.  (We note that if the same argument were applied to the
SM,
then the electron and top-quark masses should both be similar to the mass of
the SM $Z$-boson.)  Using perturbative unitarity
constraints\refmark{\pertsm,\pert}\ from the tree-level exotic fermion
scattering via \ztwo\ exchange, $F\bar F\to F\bar F$, bounds on the exotic
fermion masses can be obtained in a manner similar to the constraints obtained
on heavy fermion masses in the SM\rlap.\refmark\pertsm\
One may then ask, given the allowed
range for the exotic fermion masses, what is the likelihood that $\ztwo\to
F\bar F$ is kinematically allowed?  This probability is presented in Fig.~1
for the superstring-inspired $E_6$ effective rank-5 models (ER5M), where the
\ztwo\ couplings depend upon a parameter $-90^\circ< \theta < 90^\circ$.
In the figure, the solid curve represents the percentage of parameter space
that allows $\ztwo\to h\bar h,
E\bar E,~{\rm or}~S^c\bar S^c$, the dash-dotted curve corresponds to $\ztwo\to
N\bar N$, and the dashed curve to $\ztwo\to N^c\bar N^c$.  Note that the
probability that the \ztwo\ decays into any single pair of exotics in these
$E_6$ models is quite small, $\lsim 8\%$.  A more detailed analysis could
lead to an even smaller probability that $\ztwo\to F\bar F$ is kinematically
allowed.

We also note that the \ztwo\ production cross section into lepton pairs,
$\sigma$, also depends on the total width of the \ztwo\ and suffers some of
the same ambiguities mentioned above, although to a somewhat lesser
degree.  However, $\sigma$ can still be a valuable model discriminator.  For
example, the observation of an 8 TeV \ztwo\ {\it alone} would rule out entire
classes of models.

\chapter{EFFECTS OF DETECTOR DESCOPING}

To be as specific as possible, we will limit our descoping discussion to three
extended electroweak models:  the Left-Right Symmetric Model (LRM)\refmark\lrm\
with the ratio of right-handed to left-handed coupling constants given by
$g_R/g_L=1$, the Alternative Left-Right Model (ALRM)\rlap,\refmark\alrm\
and the Sequential Standard Model (SSM)\rlap.\refmark\ssm\  The details of
these models are also summarized in Ref.\ \us.  This particular choice was
made because these
models are fairly representative and contain no free parameters once the
\ztwo\ mass (\mtwo) is known (if $Z_1-Z_2$ mixing is neglected), and if decays
only to standard model fermions are allowed.  For numerical purposes, we
will assume an integrated luminosity (${\cal L}$) of $10\infb$ at the SSC,
corresponding to one `standard year' of run time, and take the S1 set of
Morfin-Tung parton distribution functions\refmark\mtung\ to be our canonical
set
in the calculations.

The default\refmark\sdc\ set of detector parameters that we use are:
$$
\eqalign{
&\epsilon_e = 0.85\pm 0.04 \,, \cr
&|\eta_\ell| \le 2.5 \,, \cr
&\delta M_{\ell\ell} = 0.01 M_{\ell\ell} \,, \crr
&{\delta{\cal L}\over{\cal L}} = 0.07 \,,\crr
&{\delta s\over s} = 0.10 ~{\rm with}~M_{\ell\ell}=4\tev \,, \cr} \eqno\eq
$$
where $\epsilon_e$ is the electron identification efficiency, $\eta_\ell$ is
the pseudorapidity coverage for leptons, $\delta M_{\ell\ell}$ is the mass
resolution for lepton pairs, $\delta{\cal L}/{\cal L}$ is the relative
uncertainty in the SSC integrated luminosity, and $\delta s/s$ is the relative
error in cross section and forward-backward asymmetry at $M_{\ell\ell}
=4\tev$ due to structure function uncertainties.  We note in passing that
the energy dependent term in the lepton pair mass resolution is essentially
irrelevant when dealing with new gauge bosons in the TeV mass range.

We first examine how the search limits for new gauge bosons arising from the
above three models are modified.  In setting the search limits we demand the
observation of 10 $\epem$ events arising from the \ztwo\ which are clustered
in invariant mass, with $\le 1$ event from background sources.   For the
default values of the parameters in Eq.~(2.1), the \mtwo\ discovery limits
previously obtained\refmark\us\ are 6.60 TeV (SSM), 6.10 TeV (LRM), and 6.95
TeV (ALRM).  Figures 2a-c show the percentage change in the discovery limits
as (a) the value of $\epsilon_e$, (b) the pseudorapidity cut on final state
electrons, and (c) the overall normalization of the production cross section
are altered.  These figures demonstrate that the percentage change in the
search limit is essentially model independent due only to availability of
statistics.  To confirm that there is nothing special about the three extended
models we have chosen to analyze in detail, Fig.~2a also shows the percentage
change of the search reach to modifications in $\epsilon_e$ for the
superstring-inspired $E_6$ model $\psi$ (corresponding to $\theta=
0^\circ$)\rlap.\refmark\esix\  One sees that the results obtained for this
model are very similar to the other three discussed above.  In this figure,
we see that a shift in $\epsilon_e$ of $\pm 0.10$, for example, can modify the
search reach by $-5$ to $+4\%$.  Looking at the electron pseudorapidity
dependence in
Fig.~2b, we see that (i) the discovery reach is not significantly improved when
the $\eta_\ell$ coverage is increased (since the leptons from \ztwo\ decay are
highly central) and (ii) the percentage change in the
search limit is somewhat model dependent when the $\eta_\ell$
coverage is decreased.   This is due to a modification in the lepton angular
distribution as the fermion couplings of the \ztwo\ are varied.  In summary,
while increasing $\epsilon_e$ and $\eta_\ell^{max}$ could improve the \ztwo\
search limits by at most $\simeq 4\%$, a reduction in these quantities,
if combined, could result in a $10-12\%$ decrease.  Figure 2c shows that an
uncertainty in the overall normalization of $\sigma$ does not significantly
alter the \ztwo\ discovery capability.

We now turn to the issue of model identification.  Table I shows the set
of ID-limits\refmark\us\ for the three models above (comparing two at a time),
assuming the default values of the parameters in Eq.~(2.1).  For each model in
the first column on the left, corresponding to the \ztwo\ actually
produced at the SSC, we find the maximum value of \mtwo\ for which we can
determine, at the $95\%$ CL, that the produced \ztwo\ is {\it not} from another
model.  The numbers in the Table correspond to the six possible ID-limits
that can be defined for these three distinct models.

Figures 3a-f show how the results in Table I are altered as each of the
parameters in Eq.~(2.1) are shifted from their default values.  For the
six possible pairs of models, the percentage change in the ID-limits is
presented as a function of the value of (a) $\epsilon_e$, with the uncertainty
in $\epsilon_e$ ($\delta\epsilon_e$) kept fixed at $\pm 0.04$, (b) the error in
electron identification efficiency, $\delta\epsilon_e$, with $\epsilon_e$ kept
fixed at its default value 0.85, (c) the pseudorapidity coverage for leptons,
(d) the mass resolution for lepton pairs, $\delta M_{\ell\ell}/M_{\ell\ell}$,
(e) the luminosity uncertainty $\delta\lum/\lum$, and (f) the parton
distribution uncertainty $\delta s/s$.

We see from the figures that the dependence of the ID-limits on $\epsilon_e$
is roughly model independent, as one would expect.
As $\epsilon_e$ varies by $\pm 0.10$ away from
0.85 (with $\delta\epsilon_e$ fixed), the ID-limits change at most by
$6\%$.  If $\epsilon_e$ is, however, fixed at 0.85 and $\delta
\epsilon_e$ is allowed to vary, a significant loss in the ID-limit can occur,
depending on the model, if $\delta\epsilon_e$ is poorly known.  We have also
checked that nothing is gained in the ID-limit by decreasing $\delta
\epsilon_e$ below 0.04 (for $\epsilon_e=0.85$).  As $\eta_\ell$ is varied,
we again see that nothing is gained by increasing the pseudorapidity coverage
(since the leptons are almost entirely central),
but that very substantial, albeit model dependent, losses in the ID-limits
occur if too strong a cut is made.  Unlike $\epsilon_e$, where the effect is
mainly statistical, a reduction in $\eta_\ell$ coverage not only reduces the
statistics, but also causes a reduction in the value of $A_{FB}$, which is an
important ingredient in distinguishing the \ztwo\ couplings.  Decreasing the
value of the mass resolution constant term by a factor of
3 gives at most a $\simeq 2\%$ in the ID-limits, while an increase in the mass
resolution by a factor of 2 can cost more than $\simeq 4\%$.
We see that the ID-limits are generally insensitive to variations in
$\delta\lum/\lum$ away from 0.07 with changes of at most $\pm 2\%$ as
$\delta\lum/\lum$ varies from 0.03 to 0.10.  There is a strong model
dependence in the percentage change in the ID-limits as $\delta s/s$ is
reduced below its default value of 0.10.  However, the resulting increase is
at most a few percent, even if our knowledge of structure functions in the
$M_{\ell\ell}$ range near $\simeq 4\tev$ improves by a factor of 5.

In Table II we compare the relative gains and losses in the identification
limits due to the simultaneous variation of the input parameters
to the following extreme values,
$$\eqalign{
&0.80\leq\epsilon_e\leq 0.90 \,,\cr
&0.02\leq\delta\epsilon_e\leq 0.08 \,,\cr
&2.0\leq\eta_\ell\leq 3.0 \,,\crr
&0.005\leq\delta M_{\ell\ell}/M_{\ell\ell}\leq 0.020 \,,\crr
&0.03\leq\delta\lum/\lum\leq 0.10 \,,\cr
&0.02\leq\delta s/s\leq 0.15 \,. \cr}  \eqno\eq
$$
The resulting modifications of the ID-limits are clearly quite model dependent.
However, one general feature stands out; note that the possible loss in model
determination is twice as large as the possible gain for each case.  While in
most cases there is only a modest increase in identification capability,
there is the potential for significant losses if the detector characteristics
are severely weakened.  One must keep in mind, however, that the model
identification ability is only reduced, and not destroyed altogether by
detector descoping.

\chapter{CONTRIBUTIONS FROM GAUGE BOSON MIXING}

Here we address the possible influence that mixing between the new neutral
gauge boson and the SM $Z$-boson may have on our previous
results\rlap.\refmark\xman\
To be specific, we consider the effect that such mixing has on the
determination\refmark\us\ of the parameter $\theta$ in $E_6$ models.  Recall
that the value of this parameter completely determines all fermionic couplings
of the \ztwo\ in these theories.  Defining
the weak eigenstates as $Z$ and $Z'$, the orthogonal transformation
$$\eqalign{
Z_1&=Z\cos\phi+Z'\sin\phi \,,\cr
Z_2&=-Z\sin\phi+Z'\cos\phi \,,\cr }\eqno\eq
$$
diagonalizes the $Z-Z'$ mass matrix, and produces the physical states $Z_1,Z_2$
with masses $M_1,M_2$.  $Z_1$ is then the state which is currently being
probed at LEP.  Stringent bounds on this mixing can be placed from neutral
current data, with the result\refmark\lepzp\ that $|\phi|\lsim 0.02$.

The effects of mixing are presented in Figs.\ 4a-b.  Here we show a
$\chi^2$ fit to the data on $\sigma,\ \Gamma_2,\ {\rm and}\ A_{FB}$
as a function of the $E_6$ parameter $\theta$, assuming that
a 3 TeV \ztwo\ is produced from
the two representative cases:  (a) model $\psi$ with $\theta=0^\circ$, and (b)
model $\chi$ with $\theta=-90^\circ$.  In these figures, the solid curves
represent the $\theta$ determination when the produced \ztwo\ is mixed with
the SM $Z$-boson with a value of $\phi=-0.01$ and the dashed curves
correspond to $\phi=0.01$.  Of course, since we don't know the value of
$\phi$, {\it a priori}, the fit is performed with $\phi=0$.  Thus, we are
probing the error that would be introduced in our fit to the value of
$\theta$ for case where $\phi$ is actually non-vanishing,
by assuming that $\phi=0$.
The horizontal dotted line shows the $95\%$ CL limit on the determined range
of $\theta$.  For model $\psi$ with $\phi=-0.01\ (0.01)$, the $\chi^2$ minimum
is located at $\theta=-2^\circ\ (5^\circ)$, and the $95\%$ CL determined range
of $\theta$ is $-24^\circ\ {\rm to}\ 46^\circ\ (-20^\circ\ {\rm to}\
42^\circ)$,
respectively.  Comparing these shifts to our previous results\refmark\us\
for the case where $\phi=0$ ($\chi^2$ minimum at $\theta=0^\circ$ and
$-22^\circ\leq\theta\leq 45^\circ$ at $95\%$ CL), it is clear that even
for such maximal values of mixing, there is little effect.  The contributions
from mixing are even smaller in the case of model $\chi$, where for $\phi=
-0.01\ (0.0,\ 0.01)$, the $95\%$ CL range of $\theta$ is $-114^\circ\ {\rm to}\
-66^\circ\ (-115^\circ\ {\rm to}\ -66^\circ,\ -117^\circ\ {\rm to}\
-67^\circ)$,
and the minimum is located at $\theta=-90^\circ$ for all three possible values
of $\phi$.  We conclude that our previously obtained results are robust.

\chapter{CONCLUSIONS}

In summary, we have examined the consequences of variation in detector and
machine characteristics and of neutral boson mixing in the model determination
of new neutral gauge bosons at hadron supercolliders.  We have found that
upgrades in the detector parameters over the SDC default values, do not
yield substantial improvements in \ztwo\ model differentiation, but a
severe descoping could cause appreciable
deterioration in the discovery and model identification ability.  Our results
also show that the incorporation of $Z-Z'$ mixing does not significantly
change the resulting determination of the \ztwo\ couplings.

\vskip.25in
\centerline{ACKNOWLEDGEMENTS}

This research was supported in part by awards granted by the Texas
National Research Laboratory Commission and by the U.S.~Department of Energy
under contracts W-31-109-ENG-38 and W-7405-ENG-82.

\endpage

\refout
\endpage

%
%
%
\def\vstrut{\vrule width0pt height14pt depth6pt}
\centerline{Table\ I}

\noindent{ID-limits in TeV for the various extended models discussed in
the text, assuming the default set of parameters in Eq.~(2.1).}

$$\vbox{
\offinterlineskip
\halign{
\vrule#\tabskip1em\vstrut& \hfil#\hfil& \vrule#& \hfil#\hfil& \hfil#\hfil&
\hfil#\hfil& \vrule#\tabskip0em
\cr
\noalign{\hrule height1.2pt}
width1.2pt& Produced\ \ztwo\ &width.9pt&\multispan{3}\hfil \ztwo\ Hypothesis
\hfil& width1.2pt\cr
\noalign{\vskip-6pt}
width1.2pt&  & width.9pt& SSM & LRM & ALRM & width1.2pt\cr
\noalign{\hrule height.9pt}
width1.2pt& SSM& width.9pt& --  & 5.05 & 5.95 & width1.2pt\cr
width1.2pt& LRM& width.9pt& 5.35 & --  & 6.10 & width1.2pt\cr
width1.2pt& SSM& width.9pt& 6.25 & 6.50 & --  & width1.2pt\cr
\noalign{\hrule height1.2pt} }}
$$

\endpage
%
%
%
\def\vstrut{\vrule width0pt height14pt depth6pt}
\centerline{Table\ II}

\noindent{Percentage gains and losses in ID-limits for the six pairs of models
in Table I if the input parameters are altered simultaneously.}

$$\vbox{
\offinterlineskip
\halign{
\vrule#\tabskip1em\vstrut& \hfil#\hfil& \vrule#& $\hfil#\hfil$& \vrule#&
$\hfil#\hfil$& \vrule#\tabskip0em\cr
\noalign{\hrule height1.2pt}
width1.2pt& {\rm Model} &width.9pt& {\rm Gain} (\%) && {\rm Loss}
(\%) & width1.2pt\cr
\noalign{\hrule height.9pt}
width1.2pt& LRM/ALRM & width.9pt& 2.1 && -3.1  & width1.2pt\cr
width1.2pt& LRM/SSM  & width.9pt& 5.6 && -12.4 & width1.2pt\cr
width1.2pt& ALRM/LRM & width.9pt& 3.8 && -6.6  & width1.2pt\cr
width1.2pt& ALRM/SSM & width.9pt& 5.2 && -11.3 & width1.2pt\cr
width1.2pt& SSM/LRM  & width.9pt& 9.8 && -25.6 & width1.2pt\cr
width1.2pt& SSM/ALRM & width.9pt& 5.3 && -11.8 & width1.2pt\cr
\noalign{\hrule height1.2pt} }}
$$

\endpage

%
\FIG\one{The probability that the decay $\ztwo\to F\bar F$ is kinematically
allowed for $F=h,E,S^c$ (solid curve), $F=N$ (dash-dotted curve), and
$F=N^c$ (dashed curve) in $E_6$ models.}
\FIG\two{Sensitivity to \ztwo\ discovery limits to variations in the (a)
electron identification efficiency and (b) pseudorapidity coverage for final
state leptons, for the ALRM (dashed-dotted), SSM (dashed),
LRM (solid), and the $E_6$ string-inspired model $\psi$ (dotted)}
\FIG\three{Sensitivity of the ID-limits to variations in the following detector
parameters for the six pairs of models displayed in Table I.
(a)  $\epsilon_e$ is varied with $\delta\epsilon_e$ fixed at 0.04.
{}From top to bottom, ALRM/LRM, ALRM/SSM, LRM/ALRM, SSM/ALRM, LRM/SSM, and
SSM/LRM, where the first model listed corresponds to the actual \ztwo\ that
is produced and the second to the \ztwo\ hypothesis.  (b) $\delta\epsilon_e$
is varied while $\epsilon_e$ is fixed at 0.85
for, from top to bottom, LRM/ALRM, SSM/ALRM, ALRM/SSM, ALRM/LRM,
LRM/SSM, and SSM/LRM.  (c) Percentage change in ID-limits for variations
in the $\eta_{cut}$ on leptons away from the default value.  From top to
bottom on the left-hand side of the figure, the curves are for LRM/ALRM,
LRM/SSM, ALRM/LRM, SSM/LRM, ALRM/SSM, and SSM/ALRM.  (d) Variations in the
constant term of the lepton pair mass resolution for, from top to bottom on
the right-hand side, ALRM/LRM, LRM/ALRM, LRM/SSM, SSM/LRM, ALRM/SSM, and
SSM/ALRM.  (e) Changes in the luminosity uncertainty for, from top to bottom
on the left-hand side, SSM/LRM, LRM/SSM, ALRM/SSM, ALRM/LRM, LRM/ALRM, and
\nextline
SSM/ALRM, where the last two sets of models coincide.  (f)  Alterations in
the structure function uncertainty for, from top to bottom, SSM/LRM, LRM/SSM,
ALRM/LRM, SSM/ALRM, ALRM/SSM.  Here there is no change for the set LRM/ALRM.}
\FIG\four{The $\chi^2$ determination of $\theta$, as described in the text, for
a 3 TeV \ztwo\ from $E_6$ models (a) $\psi$ ($\theta=0^\circ$) and (b) $\chi$
($\theta=-90^\circ$), including
$Z-Z'$ mixing with $\phi=-0.01(0.01)$ corresponding
to the solid (dashed) curves.}

\figout

\bye